# Choosing statistical models to assess biological interaction as a departure from additivity of effects

David M. Thompson, Yan Daniel Zhao

## Abstract

Vanderweele and Knol define biological interaction as an instance wherein "two exposures physically interact to bring about the outcome." A hallmark of biological interaction is that the total effect, produced when factors act together, differs from the sum of effects when the factors operate independently.

Epidemiologists construct statistical models to assess biological interaction. The form of the statistical model determines whether it is suited to detecting departures from additivity of effects or for detecting departures from multiplicativity of effects. A consensus exists that biological interaction should be assessed as a departure from additivity of effects.

This paper compares three statistical models' assessment of a data example that appears in several epidemiology textbooks to illustrate biological interaction in a binomial outcome. A linear binomial model quantifies departure from additivity in the data example in terms of differences in probabilities. It generates directly interpretable estimates and 95% confidence intervals for parameters including the interaction contrast (IC). Log binomial and logistic regression models detect no departure from multiplicativity in the data example. However, their estimates contribute to calculation of a "Relative Excess Risk Due to Interaction" (RERI), a measure of departure from additivity on a relative risk scale.

The linear binomial model directly produces interpretable assessments of departures from additivity of effects and deserves wider use in research and in the teaching of epidemiology. Strategies exist to address the model's limitations.

**Key Messages**

- A consensus exists in epidemiology, and is reflected in the STROBE statement, that biological interaction should be assessed as a departure from additivity of effects.
- The log binomial and logistic regression models are widely used in epidemiology to assess biological interaction, even though their statistical forms suit them for detecting departures from multiplicativity of effects.
- The log binomial and logistic regression models can quantify departures from additivity, on a relative risk scale, by estimating statistics like the Relative Excess Risk Due to Interaction (RERI). However, the RERI is not directly interpretable as excess risk, and inference (estimation and hypothesis testing) for the RERI are complicated.
- The linear model binomial model, which can be estimated using available software for generalized linear models, directly estimates the interaction contrast (IC), which is interpretable as excess risk.
- The linear binomial model deserves wider use in research and in the teaching of epidemiology.

## Biological interaction and statistical interaction

Hypotheses related to biological interaction are often of interest in studies of clinical or population health. Vanderweele and Knol [1 (p. 54)] define biological interaction as an instance in which "two exposures physically interact to bring about the outcome." Rothman [2 (p. 171)] states that "biologic interaction between two causes occurs whenever the effect of one is dependent on the presence of the other."

Investigators construct statistical models to detect interaction and effect modification. Rothman [2 (p.169)] points out that "in statistics, the term 'interaction' is used to refer to departure from the underlying form of a statistical model." A model's form can suit it for detecting departures from additivity of effects or for detecting departures from multiplicativity of effects. Because a statistical model's form affects the interpretation of statistical interaction, Rothman [2 (p.170)] prefers the term "effect measure modification" to interaction.

Rothman links "biological independence" with an additivity of effects and connects "biological interaction" with a departure from an additivity of effects. "Why is it," Rothman asks, "that biological interaction should be evaluated as departures from additivity of effect" [2 (p.178)]? By 2007, the STROBE statement regarded the response to Rothman's rhetorical question to reflect a "consensus that the additive scale, which uses absolute risks, is more appropriate [than the multiplicative scale] for public health and clinical decision making" [3 (p.817)]. The authors of the STROBE statement remind investigators that "in many circumstances, the absolute risk associated with an exposure is of greater interest than the relative risk" and ask them to "consider translating estimates of relative risk into absolute risk for a meaningful time period" [3 (p.825)]. Vanderweele and Knol [1 (p. 37)] remark, more pointedly, that "one reason why additive interaction is important to assess (rather than only relying on multiplicative interaction measures) is that it is the more relevant public health measure."

**Additivity and multiplicativity of effects**

This paper aligns with this consensus but avoids using the term "additive interaction." Instead, it links the concept to statistical models that assess evidence of a *departure from additivity of effects*. One such model, the "binomial model for the risk difference" [4], directly quantifies departures from additivity of effects in terms of differences in probabilities, including the interaction contrast (IC). This model is also called the "binomial regression model" [5, 6]. Richardson et al. [7], who employ it as a final step in a marginal structural model, call it the "linear binomial model," the term we will use.

In the linear binomial model, detection of statistical interaction constitutes direct evidence of a departure from additivity of effects. The log binomial and logistic regression models can also assess additivity indirectly, when their estimates of relative risks or odds ratios are recombined to calculate statistics like the "Relative Excess Risk due to Interaction" (RERI).

The paper also avoids using the term "multiplicative interaction" but links that concept to statistical models that assess evidence of *departures from multiplicativity of effects*. Log binomial models estimate effects in terms of relative risks, also called risk ratios, prevalence ratios [4,7] or prevalence proportion ratios. Logistic regression models estimate effects in terms of odds and odds ratios. In the log binomial and logistic models, which employ log transformations of probabilities or of their corresponding odds, detection of statistical interaction constitutes direct evidence of a departure from multiplicativity among effects.

**Statistical models for binomial outcomes**

The linear binomial, log binomial and logistic regression models are all examples of generalized linear models. Each treats the outcome as arising from a binomial distribution. Each features a linear predictor structured as a *sum of terms*. In this regard, all generalized linear models might be considered "additive." Accordingly, this paper does not refer to "additive or multiplicative models" but refers instead to statistical models that assess additivity or multiplicativity of effects.

All three models link a binomial outcome to a linear predictor. They are distinguished by the link functions they employ. The linear binomial model uses the identity link, the log binomial model uses the log link, and the logistic regression model uses the logit link. Thus, the linear binomial model operates directly on probabilities, while the others apply log transformations of the probabilities or of their corresponding odds. Because each model estimates a different effect measure, they differ in their ability to detect statistical interaction in a collection of data.

After reviewing the definition of additivity of effects, we compare the three statistical models using a widely cited example of biological interaction [8]. The linear binomial model detects statistical interaction in these data. The log binomial and logistic regression models, which assess multiplicativity of relative risks or of odds ratios, find no evidence of statistical interaction. The absence of statistical interaction in these models does not point to an absence of biological interaction, but to a lack of departure from multiplicativity of effects.

We conclude by summarizing the three models' advantages and limitations for assessing additivity of effects. The RERI is commonly used in epidemiologic research to quantify departures from additivity despite complications in its estimation, testing and interpretation. In comparison, the linear binomial model produces readily interpretable estimates of effects, including the interaction contrast.

## Defining additivity of effects

Consider a comparison of the probability or "risk" of an outcome Y among individuals who are exposed or not exposed to one or both of two "risk factors," X and Z. Then, $p_{xz}$ is a probability whose subscripts signify the probability or risk of the outcome Y at "levels" of X and Z (Table 1).

**Table 1.** Probabilities of an outcome (Y) at levels of two exposure or risk factors (X and Z)

|  | Z=1 ("exposed to factor Z") | Z=0 ("not exposed to factor Z") |
|---|---|---|
| X=1 ("exposed to factor X") | $p_{11}$ | $p_{10}$ |
| X=0 ("not exposed to factor X") | $p_{01}$ | $p_{00}$ |

Rothman [2 (p.178)] states that the following equation "establishes additivity as the definition of biological independence."

$$p_{11} - p_{00} = (p_{10} - p_{00}) + (p_{01} - p_{00}) \qquad \text{(Equation 1)}$$

According to Rothman's equation, two exposures (X and Z) are biologically independent, and their effects are additive, when the effect on Y of their joint and simultaneous effects ($p_{11} - p_{00}$) is equal to the sum of the separate and independent effects of X ($p_{10} - p_{00}$) and of Z ($p_{01} - p_{00}$). A *departure* from additivity of effect, which Rothman considers evidence of biological interaction, is present when the exposures' joint and simultaneous effect differs from the sum of their separate effects.

Additivity can be defined equivalently as a homogeneity of effects. The terms of Equation 1 can be reordered to obtain

$$p_{11} - p_{01} = p_{10} - p_{00}, \qquad \text{(Equation 2)}$$

$$p_{11} - p_{10} = p_{01} - p_{00}. \qquad \text{(Equation 3)}$$

Equation 2 states that the effect of X on Y is the same whether Z = 1 ($p_{11} - p_{01}$) or Z = 0 ($p_{10} - p_{00}$). Homogeneity of effects is reciprocal. Equation 3 states that the effect of Z on Y is the same at all levels of X, that is, whether X=1 ($p_{11} - p_{10}$) or X=0 ($p_{01} - p_{00}$). When the effects of X and Z are additive, the association between Y and X is homogenous at levels of Z, and the association between Y and Z is homogenous at levels of X.

## Assessing additivity of effects using probabilities (the interaction contrast) or ratios (the RERI)

Departures from an additivity of effects (or from biological independence), whether defined as an inequality between joint and independent effects, or as a heterogeneity among effects, can be formally assessed through the interaction contrast, whose terms are probabilities, and the RERI, whose terms are relative risks.

The terms in equation (1) can be ordered to produce the interaction contrast [9]:

$$p_{11} - p_{10} - p_{01} + p_{00} = 0 \qquad \text{(Equation 4)}$$

Reordering the terms in Equation 4 and dividing each by $p_{00}$ yields:

$$p_{11}/p_{00} - p_{01}/p_{00} - p_{10}/p_{00} + 1 = 0.$$

Recognizing that these ratios of probabilities are relative risks (RR), we obtain:

$$RR_{11} - RR_{01} - RR_{10} + 1 = 0. \qquad \text{(Equation 5)}$$

Rothman [10] names the quantity on the left side of equation 5 the "Relative Excess Risk due to Interaction" (RERI). Rothman and Greenland [9] call it the "interaction contrast ratio" (ICR).

Hosmer and Lemeshow [11] define it as "the proportion of disease among those with both exposures that is attributable to their interaction."

The algebraic equivalence between equations 1 (for the IC) and 5 (for the RERI) validates the assessment of additivity of effects on either probability or relative risk scales. The IC and the RERI formally test the hypothesis that the effects on Y of X and Z are additive or, equivalently, that no interaction exists between X and Z. The STROBE statement [3 (p.825)] illustrates how to use the RERI to assess departures from additivity of effects.

## Data example: lung cancer mortality among workers with different exposures to asbestos and smoking

Hammond et al. [8] compared the risk of a dichotomous outcome, mortality from lung cancer, among 17,800 asbestos workers and among 73,763 workers who were not exposed to asbestos. They also recorded smoking status, so participants displayed combinations of exposure to cigarette smoking and to asbestos (Table 2). Hammond's study is widely used in epidemiology textbooks [2 (pp.168-180),12] to illustrate biological interaction.

Supplementary File 1 illustrates the creation of a dataset that closely approximates the properties of the published data. So that the dataset's risk probabilities (reported as lung cancer deaths per 100,000) reflect the published ones, we assumed a smoking prevalence of 0.28 for both the asbestos workers and for the comparison group of unexposed workers.

**Table 2.** Lung cancer deaths (per 100,000 workers) among those with exposure to asbestos and/or cigarette smoking

|  | Asbestos Exposure | |
| --- | --- | --- |
| Cigarette smoking | Asbestos Workers (n= 17800) | Comparison Group (n=73763) |
| Smokers | $p_{11}$=601.9 | $p_{10}$= 121.1 |
| Non-smokers | $p_{01}$= 54.6 | $p_{00}$= 11.3 |

*The data example illustrates a departure from additivity of effects*

If the effects of asbestos exposure and cigarette smoking are additive, the expected effect of experiencing both exposures would equal the sum of the exposures' separate effects (Equation 1). Following the notation introduced in Table 1 to define $p_{xz}$, where X denotes cigarette smoking (1 = smokers and 0 = nonsmokers) and Z denotes asbestos exposure (1=exposed and 0= not exposed), the estimated risk probabilities are:

$\hat{p}_{11} - \hat{p}_{00} = 601.9 - 11.3 = 590.6$ excess deaths per 100,000 people, attributable to joint effects of both exposures.

$\hat{p}_{10} - \hat{p}_{00} = 121.0 - 11.3 = 109.7$ excess deaths per 100,000 attributable to smoking by itself.

$\hat{p}_{01} - \hat{p}_{00} = 54.6 - 11.3 = 43.3$ excess deaths per 100,000 people, attributable to asbestos exposure by itself.

The number of lung cancer deaths attributable to dual exposure appears to exceed the sum of the exposures' separate effects. The interaction contrast for the data example: $p_{11} - p_{10} - p_{01} + p_{00}$ indicates that the risk of lung cancer death in those who experience both exposures exceeds, by about 437.6 deaths per 100,000, the sum of the separate risks from smoking or from asbestos exposure. Calculated for the data example, the RERI, which quantifies additivity of effects on the relative risk scale, $RR_{11} - RR_{01} - RR_{10} + 1 = [601.9/11.3] - [54.6/11.3] - [121.0/11.3] + 1 = 38.7$.

### *The linear binomial model directly estimates the interaction contrast in the data example*

The linear binomial model [4,7] estimates the interaction contrast directly in terms of probabilities and differences in probabilities:

$$P(Y = 1) = \beta_0 + \beta_1 X + \beta_2 Z + \beta_3 XZ \qquad \text{(Equation 6)}$$

Recalling that X and Z take values of 1 for "exposure" and 0 for "no exposure", then

$$\hat{p}_{00} = \beta_0$$
$$\hat{p}_{10} - \hat{p}_{00} = (\beta_0 + \beta_1) - \beta_0 = \beta_1$$
$$\hat{p}_{01} - \hat{p}_{00} = (\beta_0 + \beta_2) - \beta_0 = \beta_2$$
$$\hat{p}_{11} - \hat{p}_{00} = (\beta_0 + \beta_1 + \beta_2 + \beta_3) - \beta_0 = \beta_1 + \beta_2 + \beta_3$$

Substituting these expressions into Equation 1, which defines additivity of effects,

$$p_{11} - p_{00} = (p_{10} - p_{00}) + (p_{01} - p_{00})$$

$$\beta_1 + \beta_2 + \beta_3 = \beta_1 + \beta_2$$

In the linear binomial model, effects are additive if $\beta_3$, the regression coefficient associated with the product or interaction term, is equal to zero.

Substituting the expressions into Equation 4 illustrates that the model's estimate for $\beta_3$ directly estimates the interaction contrast:

$$p_{11} - p_{10} - p_{01} + p_{00} = (\beta_0 + \beta_1 + \beta_2 + \beta_3) - (\beta_0 + \beta_1) - (\beta_0 + \beta_2) + \beta_0 = \beta_3$$

Thus, the linear binomial model's estimates for the interaction contrast and for the X*Z interaction are equivalent. Both provide direct tests of additivity; evidence against the hypothesis that $\beta_3 = 0$ is evidence of a departure from additivity.

Supplementary File 2 illustrates the construction of the linear binomial model using SAS PROC GENMOD [4,7]. The model's point estimates for the number of deaths per 100,000 workers, which are presented in Table 3, are equal to those reported in Table 2. Table 3 also reports the model's estimates (and 95% CI) for regression coefficients. These coefficients include estimates for the effect on lung cancer mortality of smoking among those not exposed to asbestos ($\beta_1$), and of asbestos exposure in non-smokers ($\beta_2$).

**Table 3.** Absolute risks (and risk differences) for death from lung cancer (per 100,000 workers) for those with exposure to asbestos and/or cigarette smoking, estimated by linear binomial model

|  | Smoking | Asbestos | Estimate | Deaths per 100,000 | 95% CI on estimate Lower | Upper |
|---|---|---|---|---|---|---|
| $p_{11}$ | 1 (yes) | 1 (yes) | 0.006019 | 601.926 | 387.183 | 816.669 |
| $p_{10}$ | 1 (yes) | 0 (no)  | 0.001210 | 121.048 | 73.627  | 168.469 |
| $p_{01}$ | 0 (no)  | 1 (yes) | 0.000546 | 54.619  | 14.169  | 95.070  |
| $p_{00}$ | 0 (no)  | 0 (no)  | 0.000113 | 11.298  | 2.258   | 20.337  |
| $\beta_1$ | smk ($\hat{p}_{10} - \hat{p}_{00}$) | | 0.001098 | 109.750 | 61.475 | 158.025 |
| $\beta_2$ | asbestos ($\hat{p}_{01} - \hat{p}_{00}$) | | 0.000433 | 43.322 | 1.873 | 84.770 |
| $\beta_3$ | smk*asbestos | | 0.004376 | 437.557 | 213.768 | 661.345 |
| IC | $p_{11}-p_{10}-p_{01}+p_{00}$ | | 0.004376 | 437.557 | 213.768 | 661.345 |

The linear binomial model produces identical inference for $\beta_3$, which estimates the statistical interaction between smoking and asbestos exposure, and for the IC (estimate: 437.6 deaths per 100,000; 95% CI: 213.8, 661.3; P=0.00012702). The consistency between the p values generated for these statistics verifies that they offer equivalent tests of the null hypothesis that the effects of smoking and asbestos exposure are additive.

Figure 1, which depicts the estimates and confidence intervals generated by the linear binomial model, illustrates the heterogeneity of the effects of smoking on lung cancer mortality in groups defined by asbestos exposure. The syntax that produced Table 3 and Figure 1 is contained in Supplementary File 3.

**Figure 1.** Biological interaction, between asbestos exposure and smoking, illustrated as a non-additivity or heterogeneity of effects

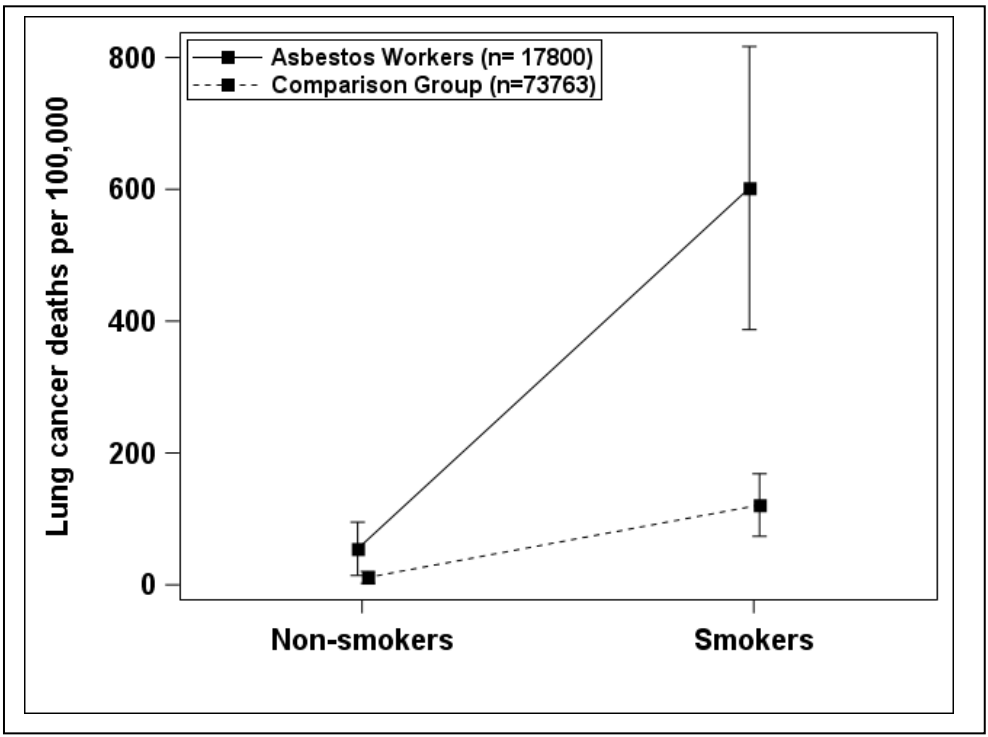

*Log binomial and logistic regression models detect no departure from multiplicativity of effects in the data example.*

In contrast to the linear binomial model, models that employ logarithmic transformations of probabilities (log binomial models) or their corresponding odds (logistic regression models) assess departures from multiplicativity of effects. Multiplicativity of effects is defined in a manner analogous to the definition of additivity of effects. The effects of two factors (X and Z) on an outcome (Y) are multiplicative if their joint effects are equal to the *product* of their separate and independent effects. When effects are multiplicative, relative risks will conform to the relationship: $RR_{XZ} = RR_X \times RR_Z$, and odds ratios will conform to the relationship: $OR_{XZ} = OR_X \times OR_Z$. A log binomial model estimates and tests the multiplicativity of relative risks.

$$\ln[P(Y = 1)] = \beta_0 + \beta_1 X + \beta_2 Z + \beta_3 XZ,$$

$$P(Y = 1) = \exp(\beta_0 + \beta_1 X + \beta_2 Z + \beta_3 XZ).$$

it follows that: $RR_{XZ} = \exp(\beta_1 X + \beta_2 Z + \beta_3 XZ)$; $RR_X = \exp(\beta_1 X)$; $RR_Z = \exp(\beta_2 Z)$.

If there is no departure from multiplicativity among relative risks, then:

$$RR_{XZ} = RR_X\, RR_Z$$

$$\exp(\beta_1 X + \beta_2 Z + \beta_3 XZ) = \exp(\beta_1 X)\exp(\beta_2 Z) = \exp(\beta_1 X + \beta_2 Z).$$

These equalities hold only if $\beta_3$, the regression coefficient associated with the product term XZ, is equal to zero. Similarly, the logistic regression model, $\ln[P(Y = 1)/P(Y = 0)] = \beta_0 + \beta_1 X + \beta_2 Z + \beta_3 XZ$, assesses multiplicativity of effects expressed as odds or odds ratios. In either model, estimates or hypothesis tests that suggest that $\beta_3$ does not equal zero constitute evidence of a departure from multiplicativity of effects.

Applied to the data example, the log binomial model finds no evidence of statistical interaction between smoking and asbestos exposure (P=0.9637); measured as relative risks, the factors' effects are multiplicative and homogenous. Similarly, a logistic regression model finds no statistical interaction between smoking and asbestos exposure (P=0.9581) to suggest a departure from multiplicativity of effects measured as odds ratios. Figures 2 and 3 depict the estimates generated by the log binomial and logistic regression models. The models' construction, using SAS PROC GENMOD, is detailed in Supplementary File 4 along with the syntax that produced Figures 2 and 3.

**Figure 2**. Predicted log probabilities illustrate a lack of departure from multiplicativity of effects in the log binomial model.

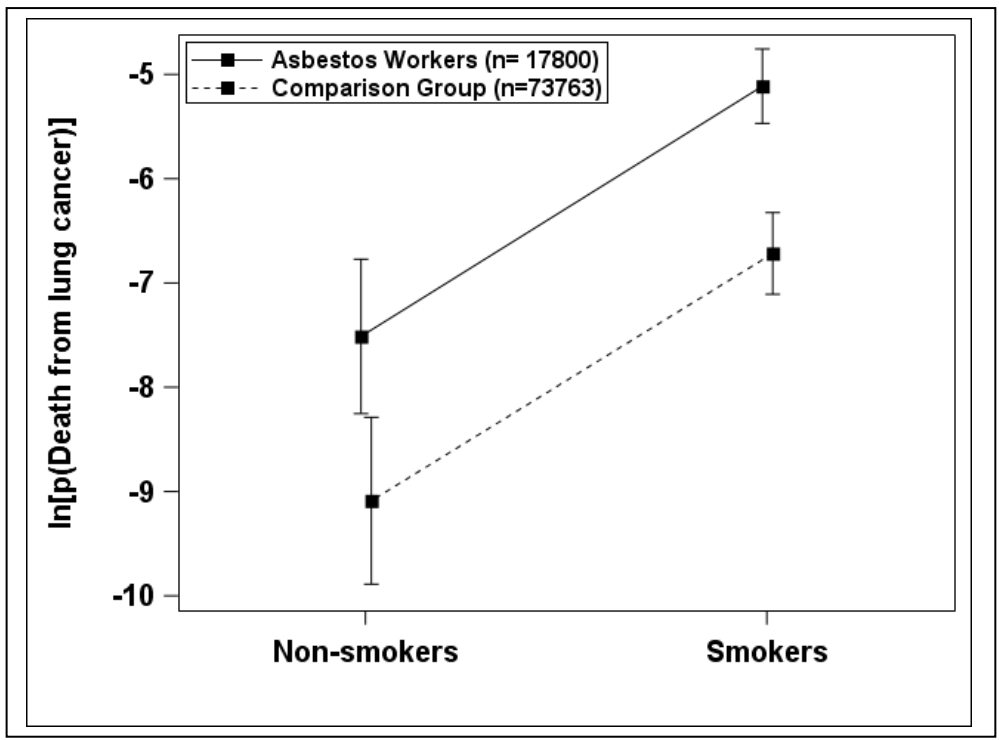

**Figure 3.** Predicted log odds illustrate a lack of departure from multiplicativity of effects in the logistic regression model.

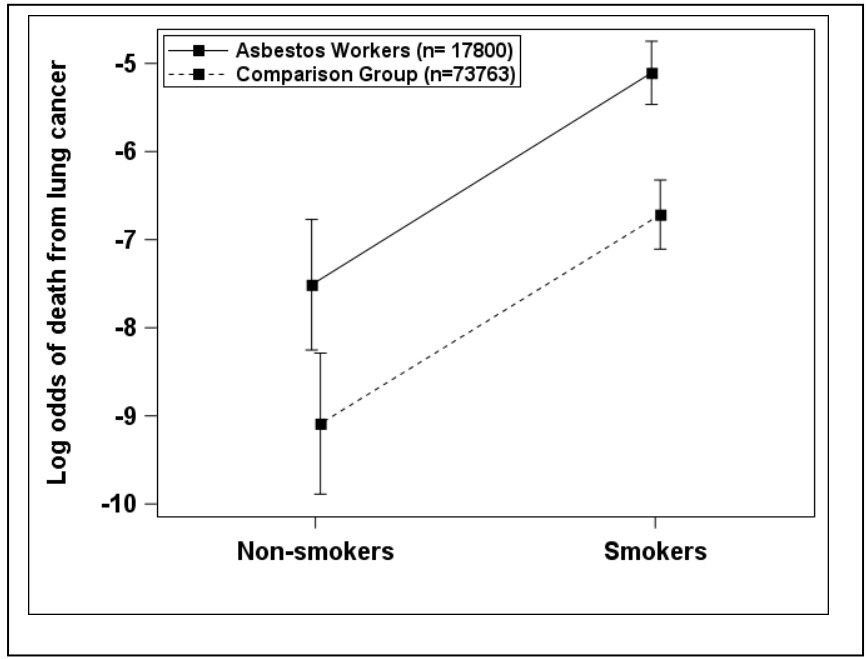

The models' differences in detecting statistical interaction do not confound the question of whether the data exemplify biological interaction. Rather, they illustrate the importance of (1) identifying an effect measure (either a difference or a ratio between probabilities or risks) that reflects the hypothesized form of the interaction and then (2) constructing a statistical model that directly estimates that effect measure.

## Choosing among statistical models

*Choosing log binomial or logistic regression models that generate estimates of the RERI*

Neither the log binomial model nor the logistic regression model detects statistical interaction in the data example. The models' form suits them for detecting departures from multiplicativity of effects. Nevertheless, they are widely used in epidemiology to assess departures from additivity of effects through ratio measures like the RERI [3].

Although widely used, the RERI has disadvantages. Because it is constructed from ratios, the RERI is not interpretable as the number of excess deaths attributable to exposure to both smoking and asbestos. The RERI of 38.7, calculated for the data example, lacks the ease of interpretation of the linear binomial model's estimate of the IC of 437.6 excess deaths per 100,000 (Table 3.) A second disadvantage relates to difficulties in obtaining standard errors with which to construct confidence intervals for or to test hypotheses related to the RERI. An influential approach, introduced by Hosmer and Lemeshow [11], estimates the RERI using logistic regression and obtains standard errors for its estimates using the delta method. SAS syntax for the approach is provided by Andersson et al. [13] and by Richardson and Kaufman [14], who construct a "linear odds ratio model" using SAS PROC NLMIXED. As an alternative approach, Richardson and Kaufmann [14] recommend bootstrapping for obtaining confidence intervals. An empirical 95% confidence interval on the RERI, calculated for the data example from 500 bootstrap samples, is 15.9, 132.6. However, because the bounds for the RERI's confidence interval are ratios, they present the same challenges to interpretation as the estimate itself.

***Choosing the linear binomial model that directly estimates the interaction contrast***

Logistic regression is widely used in epidemiology to study binomial outcomes, even though its form is suited for detecting departures from multiplicativity of effects. A major reason for the model's popularity and durability is that its use of the logit link, which is the canonical link for a binomial response, affords desirable statistical properties. Among these is logistic regression's reliability in converging on parameter estimates. Models that use other link functions can encounter problems with convergence. Zou [15] and Spiegelman and Herzmark [4] discuss problems with convergence in the log binomial model and advocate use of a modified Poisson model to address the problem when it arises.

The linear binomial model, which uses the non-canonical identity link, can also fail to converge on estimates. This limitation interferes with the model's wider acceptance, despite its ability to directly assess additivity of effects by estimating the interaction contrast. To address non-convergence in the linear binomial model, Spiegelman and Herzmark [4] advocate modifying the model, retaining the identity link but assuming that the outcome follows a Poisson distribution. Although the approach ensures convergence, imposing the Poisson assumption causes the model to misspecify the binomial outcome's variance. This intentional misspecification of the outcome's distribution reduces the efficiency of the model's standard errors and of the hypothesis tests and confidence intervals that are based on them. Accordingly, Spiegelman and Herzmark [4] recommend calculating standard errors that are robust despite misspecification. Richardson et al. [7] also recommend the calculation of robust standard errors but, because they apply it to weighted data, do not advocate otherwise modifying the linear binomial model. Supplementary File 2 shows how to incorporate these various recommendations using SAS PROC GENMOD.

Cheung [5] addresses non-convergence in the linear binomial model by proposing a modified least squares (MLS) model that also uses the identity link. Cheung's approach also calculates robust standard errors. Cheung's approach differs in that it uses ordinary least squares (OLS) instead of maximum likelihood estimation (MLE). In doing so, it avoids specifying the outcome's assumed distribution. This strategy cures the problem of non-convergence but cannot guarantee that estimated probabilities will be in the logical range from 0 to 1.

## Conclusions

Biological interaction is often hypothesized to manifest itself as a non-additivity of effects that are quantified as differences in risks or probabilities. Applied to a data example widely used in epidemiology education to illustrate biological interaction, a linear binomial model detects statistical interaction while logistic and log binomial models do not.

The result affirms the consensus that biological interaction should generally be assessed as a departure from an additivity of effects. Statistics like the RERI are widely used in epidemiology to assess additivity on a relative risk scale. In contrast, the linear binomial model produces estimates of differences in probabilities, including the interaction contrast, that are directly interpretable as excess risks.

Widely available software for generalized linear models permit researchers to construct the linear binomial model and to obtain estimates and confidence intervals for the interaction contrast and other effects. The model deserves wider use in research and judicious use in the teaching of epidemiology. The linear binomial model can encounter problems with convergence, but strategies exist to address this limitation.

**Funding**

Dr. Zhao's work was partially supported by funding provided by National Institutes of Health, National Institute of General Medical Sciences [Grant 1 U54GM104938, PI Judith James].

**Acknowledgement:**

The authors thank Dr. Tabitha Garwe for important comments on the manuscript.

**Conflict of Interest:** None declared.

**Contact information**

David M. Thompson, Department of Biostatistics and Epidemiology, University of Oklahoma Health Sciences Center, Oklahoma City, OK 73104 (e-mail: dave-thompson@ouhsc.edu).

Yan Daniel Zhao, Department of Biostatistics and Epidemiology, University of Oklahoma Health Sciences Center, Oklahoma City, OK 73104 (e-mail: daniel-zhao@ouhsc.edu).

**Supplementary File 1.** Data example

SAS syntax that creates a dataset that approximates the one published in Hammond et al. [8]. Outcome is lung cancer deaths per 100,000 workers. The prevalence of smoking is assumed to be 0.28 for both the asbestos workers and for the comparison group of unexposed workers.

```
/*Data example*/
data one;
  array asbn (2) (73763 17800); /*n in published study*/
  array rate (4) (11.3 122.6 58.4   601.6);
                   /*lung ca deaths per 100k in published study*/
  smokeprev=0.28;   /*assumed prevalence of smoking*/
  do asbestos=0 to 1;
    do smk=0 to 1;
        do lungcadeath=0 to 1;
           mult=rate [2*asbestos + smk +1] / 100000;
           count1=asbn[asbestos+1] *
                   (abs((1-smk)-smokeprev)) *
                (abs((1-lungcadeath)-mult));
           count=round(count1,1);
           output;
        end;
    end;
  end;
keep asbestos smk lungcadeath count;
run;
proc sort data=one (keep=asbestos smk lungcadeath count) out=two;
  by descending asbestos descending smk descending lungcadeath;
run;

/*version of dataset with individual observations*/
data long;
  set two;
  do i=1 to count;
    id+1;
    output;
  end;
run;

proc format;
  value smkf 1="Smokers" 0="Non-smokers";
  value gpf   1="Asbestos Workers (n= 17800)"
              0="Comparison Group (n=73763)";
  value death 1="Deaths due to lung CA"
              0="Alive or dead due to other causes";
run;

/* Table 2. Lung cancer deaths (per 100,000 workers) among those with exposure
to asbestos and/or cigarette smoking*/
proc freq data=two order=data;
  weight count;
  tables asbestos*smk*lungcadeath / nocol nopct outpct out=three;
  format smk smkf. asbestos gpf. lungcadeath death.;
run;
```

```sas
data four;
  set three;
  perhunthou=pct_row*1000;
run;

proc report nowd data=four;
  where lungcadeath=1;
  columns smk asbestos, perhunthou;
  define smk / group "Cigarette smoking" format=smkf. order=data;
  define asbestos / across "Asbestos Exposure" format=gpf. order=data;
  define perhunthou / analysis '' format=6.2;
run;
```

**Supplementary File 2.** SAS syntax for linear binomial model

The syntax below illustrates the construction of the linear binomial model [4,7] using SAS PROC GENMOD.  A MODEL statement identifies the independent variables included in comprise the linear predictor: smk (smoking status); asbestos (asbestos exposure status); and smk*asbestos, the interaction between smoking status and asbestos exposure.  Options in the MODEL statement specify that the outcome (lung cancer) follows a binomial distribution and link it directly (through an identity link) to the linear predictor.  The LSMEANS statement estimates the number of deaths per 100,000 workers for each combination of exposures.  The ESTIMATE statement lists the four coefficients $(1\ -1\ -1\ 1)$ that define the interaction contrast (IC):

$$(1)\,p_{11} + (-1)p_{10} + (-1)p_{01} + (1)p_{00} = 0$$

```
/*linear binomial model*/
proc genmod data=long descending;
  class smk (ref=first) asbestos (ref=first);
  model lungcadeath = smk asbestos smk*asbestos
          / link=identity dist=bin type3 wald;
  lsmeans smk*asbestos / cl;
  ods output lsmeans=lsmeans estimates=estimates parameterestimates=betas
modelanova=type3;
  estimate "IC" smk*asbestos 1 -1 -1 1;
run;

/*Syntax that includes a REPEATED statement, which initiates GEE estimation
of robust standard errors, advocated by Richardson et al.[7].*/
proc genmod data=long descending;
  class smk (ref=first) asbestos (ref=first) id;
  model lungcadeath = smk asbestos smk*asbestos
          / link=identity dist=bin type3 wald;
  repeated subject=id / type=ind;
  lsmeans smk*asbestos / cl;
  estimate "IC" smk*asbestos 1 -1 -1 1;
run;

/*modification of linear binomial model advocated by Spiegelman and Herzmark
[4] for instances when convergence fails*/
proc genmod data=long descending;
  class smk (ref=first) asbestos (ref=first) id;
  model lungcadeath = smk asbestos smk*asbestos
          / link=identity dist=poisson type3 wald ;
  repeated subject=id / type=ind;
  lsmeans smk*asbestos / cl;
  estimate "IC" smk*asbestos 1 -1 -1 1;
run;
```

**Supplementary File 3.** SAS syntax for Table 3 and Figure 1

The syntax below uses data sets output from the linear binomial model (Supplementary Box S2) to create Table 3 and Figure 1.

```sas
/* Table 3. Absolute risks (and risk differences) for death from lung cancer
(per 100,000 workers) for those with exposure to asbestos and/or cigarette
smoking, estimated by linear binomial model*/
data mortality;
  set lsmeans;
  mortality=estimate*100000;
  ucl=upper*100000;
  lcl=lower*100000;
run;
proc print noobs data=mortality;
  var smk asbestos estimate mortality lcl ucl;
run;

/*estimate for interaction contrast (IC)*/
data ic;
  set estimates;
  ic=meanestimate*100000;
  ic_lcl=meanlowercl*100000;
  ic_ucl=meanuppercl*100000;
run;
proc print noobs data=ic;
  var label meanestimate  ic ic_lcl ic_ucl probchisq;
  format meanestimate 9.6 probchisq 12.8 ;
run;

/*Estimates of regression coefficients, which are interpretable as excess
deaths*/
data beta2;
  set betas (where=(df=1));
  excessdeaths=estimate*100000;
  ucl=upperwaldcl*100000;
  lcl=lowerwaldcl*100000;
run;
proc print noobs data=beta2;
  var parameter estimate excessdeaths lcl ucl probchisq;
  format estimate 9.6 probchisq 12.8;
run;

/* Figure 1.  Biological interaction, between asbestos exposure and smoking,
illustrated as a non-additivity or heterogeneity of effects*/
proc template;
  define style styles.mystyle;
  parent=styles.default;
    class graphbackground / color=white;
    style GraphData1 from GraphData1 /
        contrastcolor=black linestyle=1;
    style GraphData2 from GraphData2 /
        contrastcolor=black linestyle=2;
  end;
run;
```

```
ods html style=styles.mystyle;
proc sgplot data=mortality;
   series y=mortality x=smk / group=asbestos name="one"
             groupdisplay=cluster clusterwidth=0.05
             markers markerattrs=(symbol=squarefilled size=10);
   highlow x=smk high=ucl low=lcl / group=asbestos
             groupdisplay=cluster clusterwidth=0.05
             type=line lineattrs=(pattern=1) lowcap=serif highcap=serif;
   xaxis values=(0 1) label=" " valueattrs=(size=14 weight=bold);
   yaxis label="Lung cancer deaths per 100,000"
         labelattrs=(size=14 weight=bold)
         valueattrs=(size=14 weight=bold);
   format smk smkf. asbestos gpf.;
   keylegend "one" / title="" location=inside down=2 position=topleft
             valueattrs=(size=12 weight=bold) ;
run;
ods html close;
```

**Supplementary File 4.** Log binomial and logistic regression models

Log binomial model and depiction of its estimates in Figure 2.

```
proc genmod data=long descending;
  class smk (ref=first) asbestos (ref=first)  ;
  model lungcadeath = smk asbestos smk*asbestos
        / link=log dist=bin type3 wald lrci;
  lsmeans smk*asbestos / cl;
  ods output lsmeans=lsmeans ;
run;

ods html style=styles.mystyle;
proc sgplot data=lsmeans;
  series y=estimate x=smk / group=asbestos name="one"
           groupdisplay=cluster clusterwidth=0.05
           markers markerattrs=(symbol=squarefilled size=10);
  highlow x=smk high=upper low=lower / group=asbestos
           groupdisplay=cluster clusterwidth=0.05
           type=line lineattrs=(pattern=1) lowcap=serif highcap=serif;
  xaxis values=(0 1) label=" " valueattrs=(size=14 weight=bold);
  yaxis label="ln[p(Death from lung cancer)]"
        labelattrs=(size=14 weight=bold)
        valueattrs=(size=14 weight=bold);
  format smk smkf. asbestos gpf.;
  keylegend "one" / title="" location=inside down=2 position=topleft
            valueattrs=(size=12 weight=bold) ;
run;
ods html close;
```

Logistic regression model and depiction of its estimates in Figure 3.

```
proc genmod data=long descending;
  class smk (ref=first) asbestos (ref=first)  ;
  model lungcadeath = smk asbestos smk*asbestos
        / link=logit dist=bin type3 wald lrci;
  lsmeans smk*asbestos / cl;
  ods output lsmeans=lsmeans ;
run;

ods html style=styles.mystyle;
proc sgplot data=lsmeans;
  series y=estimate x=smk / group=asbestos name="one"
           groupdisplay=cluster clusterwidth=0.05
           markers markerattrs=(symbol=squarefilled size=10);
  highlow x=smk high=upper low=lower / group=asbestos
           groupdisplay=cluster clusterwidth=0.05
           type=line lineattrs=(pattern=1) lowcap=serif highcap=serif;
  xaxis values=(0 1) label=" " valueattrs=(size=14 weight=bold);
  yaxis label="Log odds of death from lung cancer"
        labelattrs=(size=14 weight=bold)
        valueattrs=(size=14 weight=bold);
  format smk smkf. asbestos gpf.;
  keylegend "one" / title="" location=inside down=2 position=topleft
            valueattrs=(size=12 weight=bold) ;
```

```
run;
ods html close;
```